\documentstyle[preprint,aps]{revtex}
\begin{document}

\draft


\title
 {New extended  high temperature series for the\\
 $N$-vector spin models on
three-dimensional bipartite lattices\\}
\author{P. Butera and M. Comi}
\address
{Istituto Nazionale di Fisica Nucleare\\
Dipartimento di Fisica, Universit\`a di Milano\\
Via Celoria 16, 20133 Milano, Italy}
\maketitle
\begin{abstract}
Abstract: High temperature expansions
for the susceptibility and  the second correlation moment of
the classical $N$-vector model ($O(N)$ symmetric Heisenberg  model)
on the sc and the bcc lattices are extended   to  order
$\beta^{19}$ for arbitrary $N$. For
$N= 2,3,4..$ we present revised estimates of the critical parameters
 from
the newly computed coefficients.
\end{abstract}
\pacs{ PACS numbers: 05.50+q, 11.15.Ha, 64.60.Cn, 75.10.Hk}
\narrowtext

There has been a resurgence of interest in series expansions for the
Statistical Mechanics of lattice models  witnessed
 by the recent publication of several new remarkably long
high-temperature (HT)
and low-temperature (LT) series, in particular
for the $N-$vector model \cite{st68}
 with $N = 0$ (the self-avoiding walk  model)\cite{macdconw},
with $N = 1$ (the Ising model)\cite{bhanotgutent92}
and with $N = 2$ (the XY model)\cite{bcg93} in 2- and 3-space dimensions.
The best results, however, are still restricted to the
$N = 0$ and $1$ cases,
where series are   obtained by counting
techniques which achieve maximal
efficiency in low
dimensions and only
 with discrete site variable models.
Presently, on the sc lattice,  the zero field susceptibility
$\chi(N;\beta)$
and the second correlation moment $\mu_2(N;\beta)$ are known  to
 $O(\beta^{24})$
and $O(\beta^{21})$ \cite{macdconw,gut89}
respectively  for $N=0$,
    to
 $O(\beta^{19})$ and $O(\beta^{15})$\cite{gaunt,roskies}
respectively  for $N=1$,
  to  $O(\beta^{17})$ \cite{bcg93}  for $N=2$
and   to
$O(\beta^{14})$\cite{lw88,bcm90} for any other $N$.
On the bcc lattice,   $\chi$
and $\mu_2$ have been computed  to  $O(\beta^{16})$
for $N=0$ \cite{gut89},
   to $O(\beta^{21})$ for $N=1$ \cite{nickel80,nr90},
  to  $O(\beta^{12})$ for $N=2$\cite{feremore} and
  to  $O(\beta^{11})$  for $N=3$ \cite{mck}.
 Apart from  the  interest of an increasingly precise direct
determination of the critical properties of the lattice models,
there is no lack of other good reasons to undertake such a laborious
  calculation as a long series expansion:  they include
 more accurate tests  of the validity  both of the assumption of
universality,
 on which the Renormalization Group (RG)
approach to critical phenomena is based, and of
the approximation procedures
required to produce   estimates of universal quantities
by field theory methods.
In fact, for want of more rigorous arguments,
as stressed in Ref. \cite{zinn},
a  crucial test  of the validity
of Borel resummed $\epsilon - $expansions
or fixed dimension $g -$expansions
\cite{zinn,zinnbaker,murray,anton} is still provided by the comparison
with experimental or numerical data.

Here we present
a brief analysis of a new
  extension from  $O(\beta^{14})$ to $O(\beta^{19})$
 of the high temperature (HT) expansions in zero field
for the susceptibility   and the second correlation
moment   of
the $N$-vector model  both on the sc lattice and
on the bcc lattice. More results to  $O(\beta^{19})$
in $d=2,3,4,..$ space dimensions,
 and a study of the  second
field derivative of the susceptibility
 $\chi^{(4)}(N;\beta)=d^2\chi/dH^2$
will appear elsewhere.
 We have determined the series coefficients as explicit
functions of the spin dimensionality  $N$
by using  the vertex
renormalized linked cluster expansion  (LCE)
method \cite{w74mck83}. Our calculation   extends substantially
previous work in that it provides
coefficient tables of considerable length
irrespective of the spin dimensionality and of the lattice structure:
HT series  for the general
$N$-vector model were previously available only
for the  (hyper)sc
lattice in $d=2, 3, 4$ dimensions up to  $O(\beta^{14})$ \cite{lw88,bcm90}.

Concerning  the LCE technique,
we have found the following works particularly useful:
the review papers \cite{w74mck83},
 the $N=1$ computations \cite{nickel80,nr90,cd,bk,rs} and the more recent
work by M. Luescher
and P.Weisz (LW) \cite{lw88}, devoted to the  model,
the O($N$) symmetric $P(\vec \varphi^{ 2})$ lattice field theory,
described by the partition function

\begin{equation}
Z= \int \Pi d\mu(\vec \varphi_i^{ 2})
exp[ \beta \sum_{\langle i,j \rangle}\vec \varphi_i \cdot \vec \varphi_j] ,
\label{eq:uno}
\end{equation}

 where $ \vec \varphi_i$ is a $N$-component vector.
With the choice
$d\mu(\vec \varphi_i^{ 2}) = \delta(\vec \varphi_i^{ 2}-1)d\vec \varphi_i$
of the single spin
measure, (\ref{eq:uno}) reduces to the partition function of
  the $N$-vector model, but also
 a broad class of other
models of interest in Statistical Mechanics
can all be represented in this form.
LW have  devised or  simplified
some algorithms required for the calculations, and have
tabulated HT expansions of $\chi$, $\mu_2$,
$\chi^{(4)}$  on the (hyper)sc
lattices  for the $N$-vector model  to $O(\beta^{14})$\cite{lw88,bcm90}.
Also starting from (\ref{eq:uno}), we have extended the calculation
 to the class of bipartite lattices,  in particular to the
(hyper)sc and (hyper)bcc lattices. By
 redesigning the algorithms in order to take full advantage of the
structural properties of the bipartite lattices
 and  by writing an entirely new
optimized code,
we have significantly reduced the growth
of the complexity with the order of expansion.
Thus we have been able to push our calculation
well beyond
$O(\beta^{14})$ where  LW  had to give up. We can give
a rough idea of the size and the complexity
of the calculation   by mentioning
that over $2\cdot10^6$ graphs enter into the evaluation of $\chi$ and
$\mu_2$ through $O(\beta^{19})$.
This should be compared with the  corresponding
figure: $1.1\cdot10^4$, in the  LW computation.
Since these figures by no means represent our computational limits,
a further extension of our  calculations  is feasible.
We are confident that our results are correct
 also because, in each space dimension $d=1,2,3,...$,
 by a single procedure, we  produce numbers in
 agreement with all  expansion coefficients
already available for $N=$ 0,1,2,3 and $\infty$
(spherical model)\cite{nota}.
Our codes were run on
 an IBM Risc 6000/530 power
station with 32 Mbyte memory capacity and 1.5 Gbyte of disk storage.
Typical cpu times were extremely modest and
the RAM is far from being saturated.
For reasons of space neither a detailed discussion of the main
steps of this computation
can fit here, nor we can display the extensive formulas giving the
closed form
structure of the HT series coefficients as functions of $N$.
Therefore, as an example of our results,  we shall only report here the HT
series in the $N=3$ case (classical
Heisenberg model) on the sc and the bcc lattices respectively,
where we have contributed from five to eight new coefficients
beyond those given in Refs. \cite{lw88,bcm90,mck}

The susceptibility HT series for the sc lattice is:

\[
\chi^{sc}(\beta)=1 + 2 \beta + {10 \over 3}{\beta^2}
+ { 244  \over {45}} {\beta^3} +
  {230  \over {27}} {\beta^4} + {37612 \over {2835}} {\beta^5}+
  {864788 \over {42525}} {\beta^6}+
{19773464 \over {637875}} {\beta^7}+\]

\[  {89686514 \over {1913625}} {\beta^8}+
 {25478812 \over {360855}} {\beta^9}+
  {140348301868 \over {1326142125}} {\beta^{10}}+
  {477383158731608 \over {3016973334375}} {\beta^{11}}+\]

\[  {426768736125964 \over {1810184000625}} {\beta^{12}} +
  {28560817226680664 \over {81458280028125}} {\beta^{13}} +
  {775988604270248 \over {1491909890625}} {\beta^{14}} +\]

 \[ {16004552656617124832 \over {20771861407171875}} {\beta^{15}}+
  {354950851980427607594 \over {311577921107578125}} {\beta^{16}} +\]

\[  {1464128352813955096312676 \over {870237133653465703125}} {\beta^{17}} +
{1068764655864454858376417828\over {430767381158465523046875}}{\beta^{18}} +\]

\[  {259814093690188797550933157144 \over
    {71076617891146811302734375}} {\beta^{19}}...
\]

 The second correlation moment HT series for the sc lattice is:

\[
\mu_2^{sc}(\beta)= 2 \beta + 8 {\beta^2} +
  {{964 } \over {45} } {\beta^3}  +
  {{2192 } \over {45} } {\beta^4} +
  {{57116 } \over {567} } {\beta^5} +
  {{8340368 } \over {42525} } {\beta^6} +
  {{33324872 } \over {91125} } {\beta^7} +
  {{1263947744 } \over {1913625} } {\beta^8} +\]

\[  {{73478278372 } \over {63149625} } {\beta^9} +
  {{13325285538064 } \over {6630710625} } {\beta^{10}} +
  {{3434294378983784 } \over {1005657778125} } {\beta^{11} } +
  {{51819882101501984 } \over {9050920003125} } {\beta^{12} } +\]

\[  {{773005999283909656 } \over {81458280028125} } {\beta^{13} } +
  {{19031243835736702816 } \over {1221874200421875} } {\beta^{14} } +
  {{225650609227937809568 } \over {8902226317359375} } {\beta^{15} } +\]

\[  {{267912260258927725784384 } \over {6543136343259140625} } {\beta^{16} } +
  {{171544906778131647970688684 }\over{2610711400960397109375}}{\beta^{17}}+\]

\[  {{45158335170568649028207863344 } \over
    {430767381158465523046875}} {\beta^{18}} +
  {{19874973349328684680550746792 } \over
    {119456500657389598828125}} {\beta^{19}}...
\]

The susceptibility HT series for the bcc lattice is:

\[
\chi^{bcc}(\beta)=1 + {8 \over 3} \beta + {56 \over 9}{\beta^2} +
{{1936 } \over {135} } {\beta^3} +
  {{12904 } \over {405} } {\beta^4} + {{119600 } \over {1701} } {\beta^5} +
  {{2784992 } \over {18225} } {\beta^6} +
 {{632918848 } \over {1913625} } {\beta^7} +\]
\[  {{4075984504 } \over {5740875} } {\beta^8} +
  {{287925718448 } \over {189448875} } {\beta^9} +
  {{64384719769312 }  \over {19892131875} } {\beta^{10}} +
  {{186782368415874752 }  \over {27152760009375} } {\beta^{11}} +\]

\[  {{1186773786369487616 }  \over {81458280028125} } {\beta^{12}} +
  {{7528780320376815776 }  \over {244374840084375} } {\beta^{13}} +
  {{732954612970918048 }  \over {11278838773125} } {\beta^{14}} +\]

\[  {{127972570589148818590048 }  \over {934733763322734375} } {\beta^{15}} +
  {{807291210775528531339816 }  \over {2804201289968203125} } {\beta^{16}} +\]

\[ {{4736622265468109492081181616}\over{7832134202881191328125}}{\beta^{17}}+
  {{1639367056527449858924222363488 }  \over
    {1292302143475396569140625} } {\beta^{18}} +\]

\[  {{566937383305125856734568614018688  } \over
    {213229853673440433908203125} } {\beta^{19}}...
\]

 The second correlation moment HT series for the bcc lattice is:

\[
\mu_2^{bcc}(\beta)= {8  \over 3} \beta + {{128 } \over 9}{\beta^2}  +
 {{784 } \over {15} } {\beta^3} +
  {{67072 } \over {405} } {\beta^4} + {{4081648 } \over {8505} } {\beta^5} +
  {{167636864 } \over {127575} } {\beta^6} +
 {{944026304} \over {273375} } {\beta^7} +\]

\[  {{16849951744 } \over {1913625} } {\beta^8}  +
  {{4153759481008 } \over {189448875} } {\beta^9} +
  {{1065794492624896 }  \over {19892131875} } {\beta^{10}} +
  {{1166772237247486528 }  \over {9050920003125} } {\beta^{11}} +\]

\[  {{8314233519972990976 }  \over {27152760009375} } {\beta^{12}} +
  {{175799675893471696544  } \over {244374840084375} } {\beta^{13}} +
  {{409170117445661176448  } \over {244374840084375} } {\beta^{14}} +\]

\[{{3613059270200364483884384 }  \over {934733763322734375} } {\beta^{15}} +
 {{173915360520409186670373376 }\over {19629409029777421875} } {\beta^{16}} +\]

\[ {{31611058478034436738314658288}\over{1566426840576238265625}}{\beta^{17}}+
  {{8438325986405406805596333786112  } \over
    {184614591925056652734375} } {\beta^{18}} +\]

\[  {{21963940464232232523449671606261888  } \over
    {213229853673440433908203125} } {\beta^{19}}...
\]

Let us now
 comment on our updated estimates for the
critical temperatures and the critical exponents
$\gamma$ and $\nu$ in the $N=2,3,4...$ cases   where our new series
are significantly longer than those
previously available.
 The main difficulty of the analysis here comes from the expected
singular corrections\cite{wegner} (confluent singularities) to
the leading power
law behavior of thermodynamical quantities. For example, the susceptibility
 should be described, in the vicinity of its critical point $\beta_c$,
by

$ \chi(N;\beta)
\simeq C(N)(\beta_c - \beta)^{-\gamma(N)}\Big(1+ a_{\chi}(N)
(\beta_c - \beta)^{\theta(N)}+...
+ a'(N)(\beta_c - \beta)+...\Big)$

with the universal (for each N) exponent $\theta(N) \simeq 0.5$
 for small N\cite{zinn}
and $\theta(N)=  1 +O(1/N)$ for large N\cite{ma}.
The standard ratio  and Pad\'e
approximant (PA)  methods are insufficient to cope with
this very unstable  double exponential fitting problem.
Therefore  we have to resort
also to the inhomogeneous  differential
 approximants (DA) method \cite{gutasy}, a
generalization of the PA method  better
suited to represent functions behaving like $\phi_1(x) (x-x_0)^{-\gamma} +
\phi_2(x)$ near a singular point $x_0$, where $\phi_1(x)$
is a   regular function of $x$  and $\phi_2(x)$ may contain a (confluent)
singularity of
strength smaller than $\gamma$.
We have essentially followed the protocol of series
analysis by DA's  suggested
 in Ref.
\cite{gut87} which is unbiased
 for
confluent singularities. We have computed $\beta_c$ and $\gamma$
from the susceptibility series and have used this estimate of $\beta_c$
 to bias the computation of $\nu$ from the series for
 the square of the (second moment) correlation length $\xi^2$.
The  results are reported in Table I
along with the previous estimates by other
methods.
In the sc lattice case our exponent estimates are consistent with the RG
$\epsilon-$expansion results \cite{zinn,zinnbaker},
but they are  slightly larger
(by $\simeq 1\%$) than the
$g-$expansion results.
In the bcc lattice
case the estimates are perfectly compatible with the most recent seventh order
\cite{murray} or sixth order\cite{anton} $g-$expansion results.
This is  analogous to what is observed in the
most accurate unbiased analyses of the $N=1$ case\cite{gut87} and
suggests   that the series
for lattices with lower coordination number
have a  slower convergence\cite{gut87} and also
 that   unbiased DA's might be unable to account
completely for the  confluent singularities.
For $N > 3$ no elaborate estimates of the exponents by the $\epsilon-$
expansion method are available and only
very recently an extensive computation by
the (sixth order) $g-$expansion method has been published \cite{anton}.
Unfortunately, no estimates of error for the exponents are given
in Ref.\cite{anton} , but we can safely assume uncertainties of the order
of $0.5 \%$ for moderate values of $N$ and possibly smaller for $N \geq 8$.

Analysing our sc series by the simplest
biased PA method\cite{adlerholm} designed to account explicitly
for the confluent singularities,
does not significantly alter our DA estimates.
Therefore, in order to assess with a higher level
 of precision the influence of these
confluent singularities
and   completely reconcile the series results with those from
the RG,
further work is required including the computation of
even longer series and,
as indicated by the experience with the $N=1$ case,
a study of suitable continuous families of
models  for each universality class \cite{nr90,fisher}.
  The uncertainties
 we have quoted for our
 exponent estimates,     generously allowing  for  the scatter of the results
 in the DA analysis, leave
  small differences
 between our central values for the sc lattice,
and those from
the fixed dimension RG. This  suggests
that the still  insufficient length and/or the  still
incomplete account of  the confluent
singularities  add to some of our estimates a   systematic uncertainty
  twice  as large as we have indicated.
Even under this conservative proviso,
we have significantly
improved the precision of the values of the critical parameters
from HT series and  have not  pointed out any serious inconsistency
with the estimates either from
RG or from  stochastic simulations.

\acknowledgments
We are grateful to Prof. A. J. Guttmann for critically
reading a draft of this paper.

\narrowtext
\squeezetable
\begin{table}
\caption{ A summary of the estimates of critical parameters for various
values of  $N$  }
\begin{tabular}{ccccc}
$N$ & Method and Ref. & $\beta_c$ & $\gamma$&  $\nu$ \\
\hline
2& Exper.\cite{ahlers} & & & 0.6705(6)\\
 &HTE  sc &0.45420(6) & $1.328(6) $&    $0.679(3)$\\
 &HTE  bcc &0.320434(8)   &   1.323(2)&    $0.674(2)$\\
 &HTE  fcc \cite{feremore}&0.2075(1)  &    $1.323(15) $&    $0.670(7)$\\
 &R.G. g-exp. \cite{murray}& & $1.318(2)$&    $0.6715(15) $\\
 &R.G. $\epsilon$ exp.\cite{zinn}& &$1.315(7)$& $0.671(5)$\\
 &MonteCarlo sc  \cite{hasen}&0.45420(2)&     $1.308(16)$  & 0.662(7)  \\
 &MonteCarlo sc \cite{janke}&0.4542(1) &    $1.316(5) $& $0.670(7)$   \\
\hline
3 &HTE  sc &0.69302(7) & $1.403(6) $& $0.715(3)$\\
&HTE  bcc &0.48681(2)  &    1.396(3)&    $0.711(2)$\\
&HTE  fcc \cite{mck}   &0.3149(6)  &  $1.40(3) $& 0.72(1)   \\
&R.G. g-exp. \cite{murray}& & $1.3926(20)$&    $0.7096(15)$\\
&R.G. $\epsilon$ exp.\cite{zinn}& &$1.39(1)$& $0.710(7)$\\
&MonteCarlo sc \cite{chen}&0.693035(37) & $1.3896(70) $& 0.7036(23)\\
&MonteCarlo bcc \cite{chen}&0.486798(12)& $1.385(10) $& $0.7059(37)$ \\
\hline
4&HTE  sc  &0.93582(8) & $1.471(6) $&    $0.749(4)$\\
 &HTE  bcc   &0.65526(3) &   $1.458(3) $&    $0.742(2)$\\
 &R.G. g-exp.\cite{malm}& &$1.45(3)$& $0.74(1)$\\
 &R.G. g-exp.\cite{anton}& & $1.449$&    $0.738$\\
 &MonteCarlo sc \cite{kanaya} &0.9360(1) &      $1.477(18) $& 0.7479(90)\\
\hline
6&HTE  sc  &1.42838(9) & $1.577(6) $&    $0.801(4)$\\
 &HTE  bcc   &0.99608(6) &   $1.564(3) $&    $0.795(2)$\\
 &R.G. g-exp.\cite{anton}& & $1.556$&    $0.790$\\
\hline
8&HTE  sc  &1.9262(3) & $1.656(6) $&    $0.840(4)$\\
 &HTE  bcc  &1.33976(8) &   $1.641(3) $&    $0.832(2)$\\
 &R.G. g-exp. \cite{anton}& & $1.637$&    $0.830$\\
\hline
10&HTE  sc  &2.4267(3) & $1.712(6) $&    $0.867(4)$\\
 &HTE  bcc  &1.6850(2) &   $1.696(3) $&    $0.859(2)$\\
 &R.G. g-exp. \cite{anton}& & $1.697$&    $0.859$\\
\end{tabular}
\end{table}


\begin{references}
\bibitem{nota} We have corrected some very small numerical mistakes
in the last few coefficients for the $N=2$ model series
of Ref.\cite{bcg93} both in 2 and in 3 dimensions. However
the conclusions of the series analyses do not change.
\end{references}
\end{document}